# The method of solving of nonlinear Schrödinger equation

Dmitry Levko

The method of solving of nonlinear Schrödinger equation is considered. Some examples of its applications are demonstrated.

Nonlinear Schrödinger equation (NLSE)

$$i\frac{\partial \phi}{\partial t} = -\frac{\partial^2 \phi}{\partial x^2} + F(|\phi|)\cdot \phi \qquad (1)$$

arising in different physical context: in plasma's physics, in nonlinear optics and others [1]. Here $\phi = \phi(x,t)$.

The solution can be represented in the form

$$\phi(x,t) = y(z)\cdot \exp(i\delta \cdot t + ipz). \qquad (2)$$

Here $\delta$ is the real parameter, $z = x - Vt$, $p = \dfrac{V}{2}$. Than from (1)

$$y_{zz} + E_0 y - F(|y|)y = 0, \qquad (3)$$

where $E_0 = \dfrac{V^2}{4} - \delta$. The last equation we can solve by the quadratures method [2]-[3]. But we propose other method of solving (3). It look like method of solving of stationary Schrödinger equation [4].

We consider two equations

$$\begin{aligned} y'' &= (a + b\cdot y^2)y, \\ z'' &= (c + d\cdot z^2)z. \end{aligned} \qquad (4)$$

We impose such condition on solutions of (4)

$$z = A\cdot \frac{y'}{y}. \qquad (5)$$

Here $A$ is some constant (or else we get discordant correlations). If we express function $z$ through $y$ and $y'$ we get relationship between $c$ and $a$:

$$c = -2a. \qquad (6)$$

Also

$$d = 2, \qquad (7)$$

and coefficients $A$, $b$ can be arbitrary.

So if conditions (6)-(7) are carried out then both equations (4) are jointed with (5).

We shall demonstrate the method in two examples.

**(1)** The first is the next equation

$$z'' = 2\cdot(z + z^3). \qquad (8)$$

Partial solutions of this equation are known. It is

$$z_1 = \tan(x),\ \text{и}\ z_2 = \frac{1}{\tan(x)}. \qquad (9)$$

From (6) $a = -1$. We set $b = 0$ and find

$$y = A_1 \sin(x) + A_2 \cos(x). \qquad (10)$$

It follows

$$z = A\cdot \frac{A_1 \cos(x) - A_2 \sin(x)}{A_1 \sin(x) + A_2 \cos(x)}. \qquad (11)$$

If $A = -1$ and $A_1 = 0$ we derive from (11) the first of (9). If $A = 1$ and $A_2 = 0$ we derive from (11) the second of (9).

**(2)** The second is the next equation
$$z'' = -E_0 z + z^3. \tag{12}$$
We derive it from (1) where
$$F(|\phi|) = -|\phi|^2.$$
Under transformation $z \to \sqrt{2} \cdot z$ from (12) we derive
$$z'' = -E_0 z + 2z^3.$$
Then $a = \sqrt{\dfrac{E_0}{2}}$. Suppose that $b = 0$ we find
$$y = A_1 \exp(ax) + A_2 \exp(-ax),$$
and
$$z = A \cdot a \cdot \frac{A_1 \exp(ax) - A_2 \exp(-ax)}{A_1 \exp(ax) + A_2 \exp(-ax)}. \tag{13}$$
If $A_1 = A_2$ and $E_0 < 0$ equation (13) describe the black solitons [5].

Institute of Physics, National Ukrainian Academy of Science
*E-mail*: unitedlevko@yandex.ru